\documentclass[sigconf]{acmart}

\usepackage{csquotes}
\usepackage{todonotes}
\usepackage{breakcites}

\copyrightyear{2024} 
\acmYear{2024} 
\setcopyright{rightsretained} 
\acmConference[WSESE '24]{International Workshop on Methodological Issues with Empirical Studies in Software Engineering}{April 16, 2024}{Lisbon, Portugal}
\acmBooktitle{International Workshop on Methodological Issues with Empirical Studies in Software Engineering (WSESE '24), April 16, 2024, Lisbon, Portugal}\acmDOI{10.1145/3643664.3648203}
\acmISBN{979-8-4007-0567-0/24/04}

\begin{document}

\title{Evidence Tetris in the Pixelated World of Validity Threats}

\author{Marvin Wyrich}
\orcid{0000-0001-8506-3294}
\affiliation{%
  \institution{Saarland University}
  \city{Saarbrücken}
  \country{Germany}
}
\email{wyrich@cs.uni-saarland.de}

\author{Sven Apel}
\orcid{0000-0003-3687-2233}
\affiliation{%
  \institution{Saarland University}
  \city{Saarbrücken}
  \country{Germany}
}
\email{apel@cs.uni-saarland.de}

\renewcommand{\shortauthors}{Wyrich and Apel}

\begin{abstract}
    Valid empirical studies build confidence in scientific findings. Fortunately, it is now common for software engineering researchers to consider threats to validity when designing their studies and to discuss them as part of their publication.
    Yet, in complex experiments with human participants, there is often an overwhelming number of intuitively plausible threats to validity -- more than a researcher can feasibly cover.
    Therefore, prioritizing potential threats to validity becomes crucial. 
    We suggest moving away from relying solely on intuition for prioritizing validity threats, and propose that evidence on the actual impact of suspected threats to validity should complement intuition.
\end{abstract}




\maketitle

\section{Introduction}
When designing an empirical study, researchers make sure that the study design is as valid as possible. Validity is a multifaceted construct, and we will cover it in more detail later. For the moment, we use \citeauthor{Kitchenham:2015:EBSE}’s brief summary that \emph{validity refers to the degree to which we can trust the outcomes of an empirical study}~\cite{Kitchenham:2015:EBSE}.
Assessing the validity of a study design, that is, whether we can trust the results, requires expert knowledge. No one could make this assessment better than the researchers themselves, which is why we consider it a good development that nowadays threats to validity are discussed in most empirical software engineering studies~\cite{Ampatzoglou:2019:threatsSecondary,Schroeter:2017:Comprehending,Wyrich:2022:40Years,Feldt:2010:Validity}.

For example, one of the most frequently discussed threats to validity in empirical studies with human participants is the sampling of students as a substitute for professional software engineers.
To researchers who have commented publicly on this topic, the assumption that a certain treatment influences novice and expert programmers differently makes intuitive sense~\cite{Feldt:2018:FourCommentaries}.
As a result, the authors of a study often devote a paragraph in the discussion section of their paper to this potential influencing factor and mention, for example, that their sample consisted solely of students and that therefore the study results cannot be applied readily to more experienced developers.

However, most of these discussions about the validity of a study are purely based on the researcher's intuition and therefore speculative. Hardly anyone is sure about the actual extent of the discussed threats, and almost no paper cites evidence on the assumed threats. \citet{Verdecchia:2023:ThreatsCritical} describe the current situation as working through a \enquote{laundry list} of threats to validity that are only vaguely discussed.
We prefer to think of it as a pixelated world of validity threats, with no clarity about their actual impact.

This leads to uncertainty for the researchers when they design their studies, and at the very least, it leads to potential conflicts in peer review, when the reviewer critiques the study design based on their own, different intuition.
That researchers have very different views on the assessment of validity was found, for example, by \citet{Siegmund:2015:Views} in a survey of 79 program committee and editorial board members.
These views included even those that would reject papers in principle if they attempted to prioritize internal validity~\cite{Siegmund:2015:Views}.
Note that researchers should generally have their own opinions.
The fundamental issue is that personal views currently determine which scientific findings are published and which not --
a decision that, apart from ethical considerations, should instead be based on an informed validity assessment.
Otherwise, we run the risk that a few researchers will decide the scientific discourse based on their views, while valid and potentially influential minority views will be rejected unfairly.

What is needed is greater consideration of existing evidence on the influences of assumed threats to validity.
To stay with the example of student samples: whether certain characteristics of a student, such as limited programming experience, influence that person's performance and behavior in programming activities has been investigated in numerous primary studies. And we will see that the evidence does not justify speaking of a validity threat across the board in every study context.
We therefore suggest in this position paper that synthesized evidence on the most commonly assumed threats to validity can and should guide researchers in discussing the validity of their empirical software engineering study.

\section{Evidence Tetris}

We introduce the concept of \emph{Evidence Tetris} as a systematic approach for navigating validity threats in empirical  studies in software engineering research.
The approach consists of the following three steps, which build on each other and must therefore be carried out one after the other within a research field:

\begin{enumerate}
    \item \textbf{Threat collection:} Summarize the discussed threats to validity in a subfield of software engineering research.
    \item \textbf{Evidence Synthesis:} Synthesize available evidence on the most discussed threats.
    \item \textbf{Evidence-Based Study Designs:} Consider the synthesized evidence in the design and discussion of each primary study.
\end{enumerate}

The first two steps represent a community effort, on which each individual primary researcher can then build in the third step.
We describe the three steps in detail below. We illustrate the procedure and potential outcomes using the efforts of recent years within the code comprehension research community as an example.

\subsection{Threat Collection}

In everyday language, \emph{validity} can denote \enquote*{the quality of being well-grounded, sound, or correct}~\cite{mw:validity}.
In this work, validity is considered a multi-faceted construct, and the degree of validity of a study design can be assessed separately for each facet.
The facets into which validity is usually divided depend on the underlying research method of a study~\cite{Petersen:2013:Worldviews}.
In controlled experiments in software engineering, for instance, validity classifications into internal, external, construct, and conclusion have gained widespread acceptance in the reporting of validity threats.
This classification was particularly successfully promoted by \citet{Wohlin:2012:Experimentation} and Jedlitschka et al.~\cite{Jedlitschka:2008:Reporting}, but in both cases the idea is based on the work of \citet{Cook:1979:QuasiExperi}.
In principle, researchers should aim for study designs that have a high degree of validity in those facets that are most relevant to the researcher's particular principles and goals.
\enquote{Some ways of increasing one kind of validity will probably decrease another kind}~\cite{Cook:1979:QuasiExperi}, making this form of prioritization often necessary.

Yet, there still is an overwhelming number of potential threats to validity that researchers can consider and discuss.
Two secondary studies have categorized threats to validity discussed in code comprehension studies, coming up with over fifty different threat categories~\cite{Siegmund:2015:Confounding,Wyrich:2022:40Years}.
The categories cover all areas of a study design, from the selection of participants and study materials to task design, research procedure, and data analysis.

It is secondary studies such as these that make a valuable contribution within the first step of Evidence Tetris. Based on already published literature, we can find out which validity threats a research community is concerned with and which are discussed most frequently.\footnote{This can also be the result of a working group in the case of an emerging subfield.}
Furthermore, such an investigation can show what proportion of papers in a research field actually report threats to validity.
In code comprehension research, papers that do not discuss threats to validity are rather the exception today~\cite{Wyrich:2022:40Years,Schroeter:2017:Comprehending}.

\subsection{Evidence Synthesis}

Knowing which threats are discussed most frequently helps us to prioritize them in this second step of Evidence Tetris.
Our goal here is to synthesize existing evidence on validity threats. 
In an ideal world, we would synthesize evidence for all potential validity threats.
The solely pragmatic rationale for focusing on the most frequently discussed threats is that the process of evidence synthesis can be time-consuming.
Keep in mind that the most discussed threats are not necessarily the ones that have the biggest impact.
We only count their mentions to see what concerns the community the most.
Then we use evidence synthesis to see if it rightfully concerns them.

There are several ways to review literature in a secondary study \cite{Ralph:2022:Paving}.
In our context, methods that consider the influence of a threat to validity as a hypothesis and bring research findings on this hypothesis into a comprehensible and manageable structure are suitable.
These methods range from classic vote counting to statistical meta-analyses, which in extreme cases reduce research data from several studies to a single effect size.

\citet{Munoz:2023:Evidence} synthesized evidence for the three most frequently discussed threats to validity in code comprehension experiments: the influence of programming experience, program length, and the selected comprehension measures. They created \emph{evidence profiles}~\cite{Wohlin:2013:EvidenceProfile}, a form of data synthesis in which research findings are subdivided according to whether they support or refute the hypothesis, but in which each study is also assigned a qualitative strength of evidence. In an evidence profile, each individual primary study can still be recognized, which later in the third step will become advantageous to identify those studies that come closest to one's own study context.

\citet{Munoz:2023:Evidence} found that the evidence for the most frequently discussed threats in code comprehension experiments is not as clear-cut as previously assumed. It may well be that in one particular study context, for example, programming experience plays a role as a confounding variable, but in another it does not. Furthermore, for some suspected threats to validity, there is not enough evidence to even draw a conclusion. In such cases, step two of the Evidence Tetris shows where studies are still missing to explicitly investigate the significance of suspected threats.\footnote{See, e.g., \citeauthor{Wagner:2021:Intelligence}'s study~\cite{Wagner:2021:Intelligence} on the influence of intelligence and personality on code comprehension, motivated by the sole intention of investigating the actual influence of suspected confounders. Such studies provide targeted evidence for suspected threats to validity, as they make the threat an actual hypothesis of their investigation. This is not the case in all studies that provide potential evidence for suspected threats to validity.}

\subsection{Evidence-Based Study Designs}

Reaching this third and final step, a research community has already matured considerably at the meta-scientific level.
Not only has a community systematically reviewed the threats to validity that are frequently discussed within its own field of research; evidence has also been compiled for the most relevant threats to form the basis for an evidence-based discussion of study designs.
When designing and reporting a primary study, it would now be up to each researcher to consider the synthesized evidence and cite it in their reasoning for certain study design decisions.

Staying with the example of the frequently discussed influence of programming experience on the performance or behavior of developers in programming tasks. If one intuitively assumes that this influence can also play a role in one's own study, one should validate this assumption based on an already available evidence synthesis. If the assumption is confirmed, programming experience should be considered accordingly in one's own study design. In the case of a controlled experiment, this could mean that programming experience is controlled as a confounder or covariate\footnote{Ideally, a validated instrument should be used. See, e.g., the work by~\citet{Siegmund:2014:Measuring} on measuring programming experience.} (if internal validity is a concern) or that participants with a particular level of experience participate in the study (if external validity is a concern). If this cannot be accomplished, it should be discussed in the paper, citing the evidence for such a threat to validity.

In the case of a frequently suspected threat that evidently plays no role in a study context, a discussion of the threat citing other studies can also be in the researchers' own interest. The evidence then protects them from unjustified criticism.
In any case, citing evidence will contribute to a more informed validity assessment. And if authors of primary studies already cite evidence in their discussions of threats to validity, secondary studies can also benefit: contradictory findings can be better explained, and more informed and nuanced conclusions can be made.

\section{Remarks}

Each of the three steps of Evidence Tetris offer sufficient discussion material to engage in productive community conversations about the specific implementation and implications of evidence-based study designs. We would like to preface these discussions with a few remarks.

\subsection{Intuition Remains With Us}

The pleasure of science is not the least driven by the challenge of finding innovative solutions to complex problems.
Many research questions require creative and ingenious study designs, and human intuition should not be underestimated in this process.
While the goal of this paper is to bring evidence into the design and evaluation of study designs, we need to clarify two points.

First, intuition has and must have its place also in the design and discussion of future software engineering studies.
Intuition is what leads us to the hypothesis for which we collect evidence, ideas for future directions to look into, and sometimes the one methodological idea that nobody has thought about before.
Evidence will not replace intuition, but it can help a researcher defend their intuition against unwarranted criticism.

Second, the collection of evidence that this paper aims for is \emph{not} driven by a goal of coming up with the one and only true study design for a specific research question.
This line of reasoning may be intuitive, as one may at some point discern which design decisions are most logical based on empirical evidence.
And indeed, evidence will allow researchers to have meaningful discussions about their study design and how valid it is.
However, we should not forget that the validity of a study design is not binary: There is no evidence that would suggest that a study design is valid or invalid.
Evidence might tell us that a research design might be limited in some ways, and sometimes the researcher has to choose which facet of validity is important to them.
This is an intuitive process of compromise.

\subsection{Evidence and Truth}

Many years of philosophical discussions about the nature of knowledge lie behind us. If they have taught us one thing, it is that different people have different views on what evidence is and when we have enough of it to ascribe a certain temporary truth to a theory~\cite{Godfrey:2021:TheoryReality}.

For our purpose of supporting researchers in an evidence-based discussion of their empirical studies, we take a pragmatic approach and consider pieces of evidence as any information that increases our confidence in a theory.
Whether one strictly follows Popper's critical rationalism~\cite{Popper:2005:LogicDiscovery} or allows any observation to inductively contribute to this confidence in the truth of a theory, as the logical empiricists once did~\cite{Godfrey:2021:TheoryReality}, is left to the respective researcher.
Any conscious consideration of potential threats to validity in the design and reporting of studies based on past findings and observations would already represent progress compared to the current situation, in which threats are only superficially considered in a checklist-like manner~\cite{Verdecchia:2023:ThreatsCritical,Munoz:2023:Evidence}.

Regarding the question of when we have enough evidence to accept something as sufficiently supported, we do not necessarily need an answer whether something is \emph{actually} as we assume or not.
It is sufficient to know what speaks for and against to then be able to take an informed position.
This approach aligns with the thinking of Richard Rudner, who believed that evidence alone is insufficient, but that it takes a decision to accept the underlying theory~\cite{Rudner:1953:ValueJudgments}.
Thus, while we strive to draw on more evidence when evaluating study designs, we should keep these aspects in mind and, with the evidence in hand, never argue for a single absolute truth.

\subsection{The Nature of a Threats to Validity Section}

While the position presented in this paper suggests that reported threats should be supported with evidence whenever possible, this is not necessarily a sentiment shared by all members of the scientific community.
A section dedicated to validity threats may also serve as a venue for researchers to speculate without concrete evidence and identify potential deficiencies as avenues for future research.
This debate should be conducted within the respective research communities, and agreements should then be recorded and incorporated into existing guidelines for the reporting of validity threats.

We speculate that views will vary as much as they did in \citeauthor{Siegmund:2015:Views}'s survey of 79 program committee and editorial board members about their views on the importance and trade-off between internal and external validity~\cite{Siegmund:2015:Views}.
Future work could proceed and similarly investigate, for example, why researchers would (not) use evidence for the evaluation of empirical studies.
Such an investigation will potentially strike a chord and bring to light different perspectives on what constitutes relevant evidence and how much the research community may want to rely on its own past contributions.
If there are extreme views in our community, we could potentially gain insights into the causes of inconsistency in peer review decisions that have not been explored before.
And then we need to consider how to deal with the possibility that individual philosophical views may determine the publication of research findings.

\section{Conclusion}

The consideration and discussion of validity threats is not an end in itself or, worse still, like working through a laundry list.
Over the past few years, we and other researchers in the software engineering research community have recognized that there is greater potential for more informed validity assessments.

With our Evidence Tetris approach, we intend to help prioritize the overwhelming number of potential validity threats in a meaningful way.
It is a mix of community effort and the individual efforts of each researcher. 
We have illustrated that this can work using the example of the code comprehension research community, which has already successfully completed the first two of three steps.
We are confident that in the third step we will see a change in the literature towards validity discussions being more evidence-based than before.
And of course, we envision a similar momentum in other software engineering research communities.

\section*{Acknowledgments}
We would like to thank three anonymous reviewers for their constructive criticism.
Wyrich's and Apel's work is supported by the European Union as part of the ERC Advanced Grant 101052182. Parts of this paper are drawn from Wyrich's doctoral thesis~\cite{Wyrich:2023:Evidence}.

\bibliographystyle{ACM-Reference-Format}
\bibliography{references}


\begin{thebibliography}{23}


\ifx \showCODEN    \undefined \def \showCODEN     #1{\unskip}     \fi
\ifx \showDOI      \undefined \def \showDOI       #1{#1}\fi
\ifx \showISBNx    \undefined \def \showISBNx     #1{\unskip}     \fi
\ifx \showISBNxiii \undefined \def \showISBNxiii  #1{\unskip}     \fi
\ifx \showISSN     \undefined \def \showISSN      #1{\unskip}     \fi
\ifx \showLCCN     \undefined \def \showLCCN      #1{\unskip}     \fi
\ifx \shownote     \undefined \def \shownote      #1{#1}          \fi
\ifx \showarticletitle \undefined \def \showarticletitle #1{#1}   \fi
\ifx \showURL      \undefined \def \showURL       {\relax}        \fi
\providecommand\bibfield[2]{#2}
\providecommand\bibinfo[2]{#2}
\providecommand\natexlab[1]{#1}
\providecommand\showeprint[2][]{arXiv:#2}

\bibitem[Ampatzoglou et~al\mbox{.}(2019)]%
        {Ampatzoglou:2019:threatsSecondary}
\bibfield{author}{\bibinfo{person}{Apostolos Ampatzoglou}, \bibinfo{person}{Stamatia Bibi}, \bibinfo{person}{Paris Avgeriou}, \bibinfo{person}{Marijn Verbeek}, {and} \bibinfo{person}{Alexander Chatzigeorgiou}.} \bibinfo{year}{2019}\natexlab{}.
\newblock \showarticletitle{Identifying, categorizing and mitigating threats to validity in software engineering secondary studies}.
\newblock \bibinfo{journal}{\emph{Information and Software Technology}}  \bibinfo{volume}{106} (\bibinfo{year}{2019}), \bibinfo{pages}{201--230}.
\newblock
\showISSN{0950-5849}
\urldef\tempurl%
\url{https://doi.org/10.1016/j.infsof.2018.10.006}
\showDOI{\tempurl}


\bibitem[Cook and Campbell(1979)]%
        {Cook:1979:QuasiExperi}
\bibfield{author}{\bibinfo{person}{Thomas~D Cook} {and} \bibinfo{person}{Donald~Thomas Campbell}.} \bibinfo{year}{1979}\natexlab{}.
\newblock \bibinfo{booktitle}{\emph{Quasi-experimentation: Design \& analysis issues for field settings}}. Vol.~\bibinfo{volume}{351}.
\newblock \bibinfo{publisher}{Houghton Mifflin Boston}.
\newblock
\showISBNx{0528620533}


\bibitem[Feldt and Magazinius(2010)]%
        {Feldt:2010:Validity}
\bibfield{author}{\bibinfo{person}{Robert Feldt} {and} \bibinfo{person}{Ana Magazinius}.} \bibinfo{year}{2010}\natexlab{}.
\newblock \showarticletitle{Validity threats in empirical software engineering research-an initial survey.}. In \bibinfo{booktitle}{\emph{Seke}}. \bibinfo{pages}{374--379}.
\newblock


\bibitem[Feldt et~al\mbox{.}(2018)]%
        {Feldt:2018:FourCommentaries}
\bibfield{author}{\bibinfo{person}{Robert Feldt}, \bibinfo{person}{Thomas Zimmermann}, \bibinfo{person}{Gunnar~R Bergersen}, \bibinfo{person}{Davide Falessi}, \bibinfo{person}{Andreas Jedlitschka}, \bibinfo{person}{Natalia Juristo}, \bibinfo{person}{J{\"u}rgen M{\"u}nch}, \bibinfo{person}{Markku Oivo}, \bibinfo{person}{Per Runeson}, \bibinfo{person}{Martin Shepperd}, {et~al\mbox{.}}} \bibinfo{year}{2018}\natexlab{}.
\newblock \showarticletitle{Four commentaries on the use of students and professionals in empirical software engineering experiments}.
\newblock \bibinfo{journal}{\emph{Empirical Software Engineering}} \bibinfo{volume}{23}, \bibinfo{number}{6} (\bibinfo{year}{2018}), \bibinfo{pages}{3801--3820}.
\newblock


\bibitem[Godfrey-Smith(2021)]%
        {Godfrey:2021:TheoryReality}
\bibfield{author}{\bibinfo{person}{Peter Godfrey-Smith}.} \bibinfo{year}{2021}\natexlab{}.
\newblock \bibinfo{booktitle}{\emph{Theory and Reality: An Introduction to the Philosophy of Science, Second Edition}}.
\newblock \bibinfo{publisher}{University of Chicago Press}.
\newblock
\showISBNx{9780226771137}


\bibitem[Jedlitschka et~al\mbox{.}(2008)]%
        {Jedlitschka:2008:Reporting}
\bibfield{author}{\bibinfo{person}{Andreas Jedlitschka}, \bibinfo{person}{Marcus Ciolkowski}, {and} \bibinfo{person}{Dietmar Pfahl}.} \bibinfo{year}{2008}\natexlab{}.
\newblock \showarticletitle{Reporting experiments in software engineering}.
\newblock In \bibinfo{booktitle}{\emph{Guide to advanced empirical software engineering}}. \bibinfo{publisher}{Springer}, \bibinfo{pages}{201--228}.
\newblock


\bibitem[Kitchenham et~al\mbox{.}(2015)]%
        {Kitchenham:2015:EBSE}
\bibfield{author}{\bibinfo{person}{Barbara~Ann Kitchenham}, \bibinfo{person}{David Budgen}, {and} \bibinfo{person}{Pearl Brereton}.} \bibinfo{year}{2015}\natexlab{}.
\newblock \bibinfo{booktitle}{\emph{Evidence-Based Software Engineering and Systematic Reviews}}. \bibinfo{series}{Chapman {\&} Hall / CRC Innovations in Software Engineering and Software Development Series}, Vol.~\bibinfo{volume}{4}.
\newblock \bibinfo{publisher}{{CRC Press}}, \bibinfo{address}{Boca Raton}.
\newblock
\showISBNx{978-1-4822-2865-6}


\bibitem[Merriam-Webster({[n.\,d.]})]%
        {mw:validity}
\bibfield{author}{\bibinfo{person}{Merriam-Webster}.} \bibinfo{year}{[n.\,d.]}\natexlab{}.
\newblock \showarticletitle{Validity}.
\newblock In \bibinfo{booktitle}{\emph{Merriam-Webster.com dictionary}}.
\newblock
\urldef\tempurl%
\url{https://www.merriam-webster.com/dictionary/validity}
\showURL{%
\tempurl}


\bibitem[Mu{\~n}oz~Bar{\'o}n et~al\mbox{.}(2023)]%
        {Munoz:2023:Evidence}
\bibfield{author}{\bibinfo{person}{Marvin Mu{\~n}oz~Bar{\'o}n}, \bibinfo{person}{Marvin Wyrich}, \bibinfo{person}{Daniel Graziotin}, {and} \bibinfo{person}{Stefan Wagner}.} \bibinfo{year}{2023}\natexlab{}.
\newblock \showarticletitle{Evidence Profiles for Validity Threats in Program Comprehension Experiments}. In \bibinfo{booktitle}{\emph{2023 IEEE/ACM 45th International Conference on Software Engineering (ICSE)}}. IEEE, \bibinfo{publisher}{IEEE}.
\newblock


\bibitem[Petersen and Gencel(2013)]%
        {Petersen:2013:Worldviews}
\bibfield{author}{\bibinfo{person}{Kai Petersen} {and} \bibinfo{person}{Cigdem Gencel}.} \bibinfo{year}{2013}\natexlab{}.
\newblock \showarticletitle{Worldviews, Research Methods, and their Relationship to Validity in Empirical Software Engineering Research}. In \bibinfo{booktitle}{\emph{2013 Joint Conference of the 23rd International Workshop on Software Measurement and the 8th International Conference on Software Process and Product Measurement}}. \bibinfo{pages}{81--89}.
\newblock
\urldef\tempurl%
\url{https://doi.org/10.1109/IWSM-Mensura.2013.22}
\showDOI{\tempurl}


\bibitem[Popper(2005)]%
        {Popper:2005:LogicDiscovery}
\bibfield{author}{\bibinfo{person}{K. Popper}.} \bibinfo{year}{2005}\natexlab{}.
\newblock \bibinfo{booktitle}{\emph{The Logic of Scientific Discovery}}.
\newblock \bibinfo{publisher}{Taylor \& Francis}.
\newblock
\showISBNx{9781134470020}


\bibitem[Ralph and Baltes(2022)]%
        {Ralph:2022:Paving}
\bibfield{author}{\bibinfo{person}{Paul Ralph} {and} \bibinfo{person}{Sebastian Baltes}.} \bibinfo{year}{2022}\natexlab{}.
\newblock \showarticletitle{Paving the Way for Mature Secondary Research: The Seven Types of Literature Review}. In \bibinfo{booktitle}{\emph{Proceedings of the 30th ACM Joint European Software Engineering Conference and Symposium on the Foundations of Software Engineering}} (Singapore) \emph{(\bibinfo{series}{ESEC/FSE 2022})}. \bibinfo{publisher}{Association for Computing Machinery}, \bibinfo{address}{New York, NY, USA}, \bibinfo{pages}{1632–1636}.
\newblock
\showISBNx{9781450394130}
\urldef\tempurl%
\url{https://doi.org/10.1145/3540250.3560877}
\showDOI{\tempurl}


\bibitem[Rudner(1953)]%
        {Rudner:1953:ValueJudgments}
\bibfield{author}{\bibinfo{person}{Richard Rudner}.} \bibinfo{year}{1953}\natexlab{}.
\newblock \showarticletitle{The Scientist Qua Scientist Makes Value Judgments}.
\newblock \bibinfo{journal}{\emph{Philosophy of Science}} \bibinfo{volume}{20}, \bibinfo{number}{1} (\bibinfo{year}{1953}), \bibinfo{pages}{1--6}.
\newblock
\showISSN{00318248, 1539767X}


\bibitem[Schröter et~al\mbox{.}(2017)]%
        {Schroeter:2017:Comprehending}
\bibfield{author}{\bibinfo{person}{Ivonne Schröter}, \bibinfo{person}{Jacob Krüger}, \bibinfo{person}{Janet Siegmund}, {and} \bibinfo{person}{Thomas Leich}.} \bibinfo{year}{2017}\natexlab{}.
\newblock \showarticletitle{Comprehending Studies on Program Comprehension}. In \bibinfo{booktitle}{\emph{2017 IEEE/ACM 25th International Conference on Program Comprehension (ICPC)}}. \bibinfo{pages}{308--311}.
\newblock
\urldef\tempurl%
\url{https://doi.org/10.1109/ICPC.2017.9}
\showDOI{\tempurl}


\bibitem[Siegmund et~al\mbox{.}(2014)]%
        {Siegmund:2014:Measuring}
\bibfield{author}{\bibinfo{person}{Janet Siegmund}, \bibinfo{person}{Christian K{\"a}stner}, \bibinfo{person}{J{\"o}rg Liebig}, \bibinfo{person}{Sven Apel}, {and} \bibinfo{person}{Stefan Hanenberg}.} \bibinfo{year}{2014}\natexlab{}.
\newblock \showarticletitle{Measuring and modeling programming experience}.
\newblock \bibinfo{journal}{\emph{Empirical Software Engineering}}  \bibinfo{volume}{19} (\bibinfo{year}{2014}), \bibinfo{pages}{1299--1334}.
\newblock


\bibitem[Siegmund and Schumann(2015)]%
        {Siegmund:2015:Confounding}
\bibfield{author}{\bibinfo{person}{Janet Siegmund} {and} \bibinfo{person}{Jana Schumann}.} \bibinfo{year}{2015}\natexlab{}.
\newblock \showarticletitle{Confounding parameters on program comprehension: a literature survey}.
\newblock \bibinfo{journal}{\emph{Empirical Software Engineering}} \bibinfo{volume}{20}, \bibinfo{number}{4} (\bibinfo{year}{2015}), \bibinfo{pages}{1159--1192}.
\newblock


\bibitem[Siegmund et~al\mbox{.}(2015)]%
        {Siegmund:2015:Views}
\bibfield{author}{\bibinfo{person}{Janet Siegmund}, \bibinfo{person}{Norbert Siegmund}, {and} \bibinfo{person}{Sven Apel}.} \bibinfo{year}{2015}\natexlab{}.
\newblock \showarticletitle{Views on internal and external validity in empirical software engineering}. In \bibinfo{booktitle}{\emph{2015 IEEE/ACM 37th IEEE International Conference on Software Engineering}}, Vol.~\bibinfo{volume}{1}. IEEE, \bibinfo{publisher}{IEEE}, \bibinfo{pages}{9--19}.
\newblock


\bibitem[Verdecchia et~al\mbox{.}(2023)]%
        {Verdecchia:2023:ThreatsCritical}
\bibfield{author}{\bibinfo{person}{Roberto Verdecchia}, \bibinfo{person}{Emelie Engström}, \bibinfo{person}{Patricia Lago}, \bibinfo{person}{Per Runeson}, {and} \bibinfo{person}{Qunying Song}.} \bibinfo{year}{2023}\natexlab{}.
\newblock \showarticletitle{Threats to validity in software engineering research: A critical reflection}.
\newblock \bibinfo{journal}{\emph{Information and Software Technology}}  \bibinfo{volume}{164} (\bibinfo{year}{2023}), \bibinfo{pages}{107329}.
\newblock
\showISSN{0950-5849}
\urldef\tempurl%
\url{https://doi.org/10.1016/j.infsof.2023.107329}
\showDOI{\tempurl}


\bibitem[Wagner and Wyrich(2021)]%
        {Wagner:2021:Intelligence}
\bibfield{author}{\bibinfo{person}{Stefan Wagner} {and} \bibinfo{person}{Marvin Wyrich}.} \bibinfo{year}{2021}\natexlab{}.
\newblock \showarticletitle{Code Comprehension Confounders: A Study of Intelligence and Personality}.
\newblock \bibinfo{journal}{\emph{IEEE Transactions on Software Engineering}} \bibinfo{volume}{48}, \bibinfo{number}{12} (\bibinfo{year}{2021}), \bibinfo{pages}{4789--4801}.
\newblock


\bibitem[Wohlin(2013)]%
        {Wohlin:2013:EvidenceProfile}
\bibfield{author}{\bibinfo{person}{Claes Wohlin}.} \bibinfo{year}{2013}\natexlab{}.
\newblock \bibinfo{booktitle}{\emph{An Evidence Profile for Software Engineering Research and Practice}}.
\newblock \bibinfo{publisher}{Springer Berlin Heidelberg}, \bibinfo{address}{Berlin, Heidelberg}, \bibinfo{pages}{145--157}.
\newblock
\showISBNx{978-3-642-37395-4}
\urldef\tempurl%
\url{https://doi.org/10.1007/978-3-642-37395-4_10}
\showDOI{\tempurl}


\bibitem[Wohlin et~al\mbox{.}(2012)]%
        {Wohlin:2012:Experimentation}
\bibfield{author}{\bibinfo{person}{Claes Wohlin}, \bibinfo{person}{Per Runeson}, \bibinfo{person}{Martin H{\"{o}}st}, \bibinfo{person}{Magnus~C. Ohlsson}, \bibinfo{person}{Bj{\"{o}}rn Regnell}, {and} \bibinfo{person}{Anders Wessl{\'{e}}n}.} \bibinfo{year}{2012}\natexlab{}.
\newblock \bibinfo{booktitle}{\emph{{Experimentation in Software Engineering}}}. Vol.~\bibinfo{volume}{9783642290}.
\newblock \bibinfo{publisher}{Springer Berlin Heidelberg}, \bibinfo{address}{Berlin, Heidelberg}. 1--236 pages.
\newblock
\showISBNx{978-3-642-29043-5}
\urldef\tempurl%
\url{https://doi.org/10.1007/978-3-642-29044-2}
\showDOI{\tempurl}


\bibitem[Wyrich(2023)]%
        {Wyrich:2023:Evidence}
\bibfield{author}{\bibinfo{person}{Marvin Wyrich}.} \bibinfo{year}{2023}\natexlab{}.
\newblock \emph{\bibinfo{title}{Evidence for the design of code comprehension experiments}}.
\newblock \bibinfo{thesistype}{Ph.\,D. Dissertation}. \bibinfo{school}{University of Stuttgart}.
\newblock


\bibitem[Wyrich et~al\mbox{.}(2023)]%
        {Wyrich:2022:40Years}
\bibfield{author}{\bibinfo{person}{Marvin Wyrich}, \bibinfo{person}{Justus Bogner}, {and} \bibinfo{person}{Stefan Wagner}.} \bibinfo{year}{2023}\natexlab{}.
\newblock \showarticletitle{40 Years of Designing Code Comprehension Experiments: A Systematic Mapping Study}.
\newblock \bibinfo{journal}{\emph{ACM Comput. Surv.}} (\bibinfo{year}{2023}).
\newblock
\showISSN{0360-0300}
\urldef\tempurl%
\url{https://doi.org/10.1145/3626522}
\showDOI{\tempurl}


\end{thebibliography}

\end{document}